\documentclass[prb,preprint,titlepage,amsmath,amssymb,superscriptaddress,longbibliography]{revtex4-1}

\usepackage[utf8]{inputenc}
\usepackage[english]{babel}
\usepackage[babel,german=quotes]{csquotes}
\usepackage[T1]{fontenc}
\usepackage{amsmath}
\usepackage{amssymb}
\usepackage{graphicx}

\usepackage{soul}

\usepackage{xcolor}

\DeclareMathOperator{\sech}{sech}

\newcommand{\normord}[1]{%
{:\mathrel{\mspace{1mu}#1\mspace{1mu}}:}%
}

\definecolor{mlgreen}{rgb}{.035,.6,.251}
\definecolor{mlviolett}{rgb}{.643,.259,.804}

\begin{document}

\title{Subcycle squeezing of light from a time flow perspective}
\author{Matthias Kizmann}
\affiliation{Department of Physics and Center for Applied Photonics,
University of Konstanz, D-78457 Konstanz, Germany}
\author{Thiago Lucena de M. Guedes}
\affiliation{Department of Physics and Center for Applied Photonics,
University of Konstanz, D-78457 Konstanz, Germany}
\author{Denis V. Seletskiy}
\affiliation{Department of Engineering Physics, Polytechnique Montréal, H3T 1J4, Canada}
\affiliation{Department of Physics and Center for Applied Photonics,
University of Konstanz, D-78457 Konstanz, Germany}
 \author{Andrey S. Moskalenko}
 \email{andrey.moskalenko@uni-konstanz.de}
 \affiliation{Department of Physics and Center for Applied Photonics,
University of Konstanz, D-78457 Konstanz, Germany}
 \author{Alfred Leitenstorfer}
 \affiliation{Department of Physics and Center for Applied Photonics,
University of Konstanz, D-78457 Konstanz, Germany}
  \author{Guido Burkard}
  \email{guido.burkard@uni-konstanz.de}
  \affiliation{Department of Physics and Center for Applied Photonics,
University of Konstanz, D-78457 Konstanz, Germany}

\maketitle

\noindent
\textbf{Light as a carrier of information and energy plays a fundamental role in both general relativity and quantum physics, linking these areas that are still not fully compliant with each other. Its quantum nature and spatio-temporal structure are exploited in many intriguing applications ranging from novel spectroscopy methods of complex many-body phenomena \cite{Dorfman2016}  to quantum information processing \cite{Knill2001,Weedbrook2012,Broome2013} and subwavelength lithography \cite{Boto2000,DAngelo2001}. Recent access to subcycle quantum features of electromagnetic radiation \cite{Riek2015,Moskalenko2015,Benea2016,Riek2017} promises a new class of time-dependent quantum states of light. Paralleled with the developments in attosecond science \cite{Corkum2007,Krausz2009,Moskalenko_Phys_Rep2017}, these advances motivate an urgent need for a theoretical framework that treats arbitrary wave packets of quantum light intrinsically in the time domain.
Here, we formulate a consistent time domain theory of the generation and sampling of few-cycle and subcycle pulsed squeezed states, allowing for a relativistic interpretation in terms of induced changes in the local flow of time. Our theory enables the use of such states as a resource for novel ultrafast applications in quantum optics and quantum information. 
}

In conventional optics, there was no control over the absolute phase of a light pulse with respect to its envelope. This situation changed abruptly with the advent of femtosecond frequency combs\cite{Jones2000,Holzwarth2000}. The carrier-envelope phase is especially relevant for few-cycle pulses which are at the heart of attosecond technology and extremely nonlinear optics\cite{Corkum2007,Krausz2009}. So far, this area has exploited coherent states of light which come closest to the classical picture of an electromagnetic field with well-defined amplitude and phase. Only recently, quantum optics was carried to an analogous level with subcycle analysis of the noise properties of infrared electric fields by electro-optic sampling with femtosecond laser pulses \cite{Riek2015,Moskalenko2015,Benea2016,Riek2017}. Here, the concept of a carrier-envelope phase loses its meaning because highly nonclassical states of light may exhibit excessive phase fluctuations or no well-defined phase at all. Instead, it is the relative timing of a quantum noise pattern with respect to a subcycle probe which gains relevance. 
Interesting and unexpected phenomena may arise when the physics of such synchronal states of light is explored on a subcycle scale.

One of
the most fundamental nonclassical states of light is the squeezed vacuum state \cite{Breitenbach1997}. Employed in an interferometer, it can provide a detector sensitivity below the shot-noise limit \cite{Caves1981} that was used in the measurements leading to the recent experimental discovery of gravitational waves\cite{Abbott2016}. 
Stationary squeezed vacuum states  carry a persisting flux of photons \cite{Blow1990}, whereas there is a finite number of photons contained in a pulsed state. Pulsed squeezed light \cite{Anderson1997} with many optical oscillation cycles was realized based on parametric down-conversion \cite{Slusher1987}. The case of the generation of pulsed broadband ultrashort squeezing\cite{Wasilewski2006,Harris2007,Horoshko2013} was analyzed in terms of a large set of single-frequency or broadband shaped temporal modes \cite{Wasilewski2006,Christ2013,Peer2018}, where each mode is characterized
separately.
 The alternative time-domain
approach based on electro-optic sampling provided first insights into the temporally
resolved dynamics of few-cycle squeezing supported by a simplified theoretical description \cite{Riek2017}.
This theory was restricted to the low squeezing regime or selected points in time. Furthermore, it has not yet incorporated a finite probe pulse duration for the detection of the temporal quantum noise patterns. Here, we present methods that overcome these limitations by considering extremely short (in terms of optical cycles) pulses of squeezed vacuum light and provide a general theoretical picture of their generation and subcycle-resolved detection. We predict that the expected asymmetry between the anti-squeezed and squeezed temporal noise, well known in conventional quantum optics, can in fact be reversed for ultrabroadband driving fields which are also within reach of state-of-the-art experiments.  Our results shed light on the interplay between the quantum nature of electromagnetic fields and general relativity, leading to creation of particles out of vacuum on ultrashort time scales.

\begin{figure}[t!]
\includegraphics[scale=0.4]{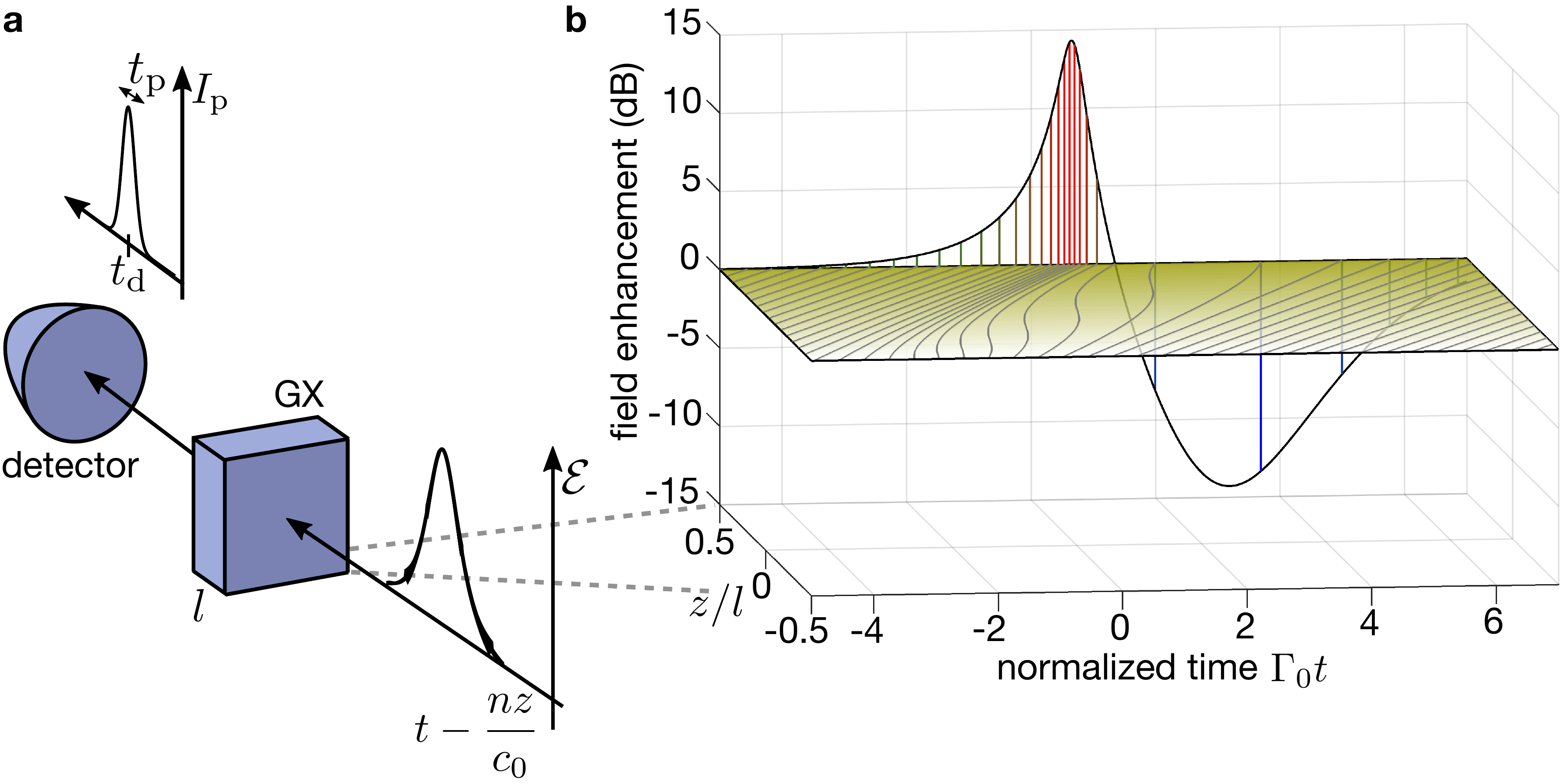}
\caption{\textbf{Scheme
of the generation and detection setup and the corresponding evolution of the mid-infrared quantum field $\hat{\varepsilon}(z,t)$ inside the nonlinear crystal for a half-cycle mid-infrared driving field with effective squeezing strength $r=5$. a}, A strong mid-infrared coherent field $\mathcal{E}(z,t)$ is sent into the nonlinear generating crystal (GX) of length $l$ where it squeezes the co-propagating vacuum field $\hat{\varepsilon}(z,t)$. The squeezed quantum field is then detected using electro-optic sampling: a probe pulse with intensity envelope $I_\mathrm{p}$ plays the role of a temporal gating applied at various time points $t_\mathrm{d}$. Statistical readout allows to obtain the time-resolved variance of $\hat{\varepsilon}(z,t)$.
Subcycle resolution is achieved for sufficiently short probe pulse durations $t_\mathrm{p}$. \textbf{b},
Horizontal plane: Grey lines depict the world lines for $\hat{\varepsilon}(z,t)$  [determined by $\tau(z,t)=\mathrm{const}$, characteristic curves of Eq.~\eqref{PDE}], shown for the case of a half-cycle driving field \eqref{HCP} with duration $\Gamma_0^{-1}$. The equidistant spacing of the lines at
the entrance gets distorted as the quantum light propagates through the crystal. Vertical plane: Field enhancement at the crystal exit is shown
together with the final spacing between the world lines. In the simplified picture, red (blue) lines correspond to anti-squeezing (squeezing). \label{Fig0}}
\end{figure}
\begin{figure}[t!]
\includegraphics[scale=0.4]{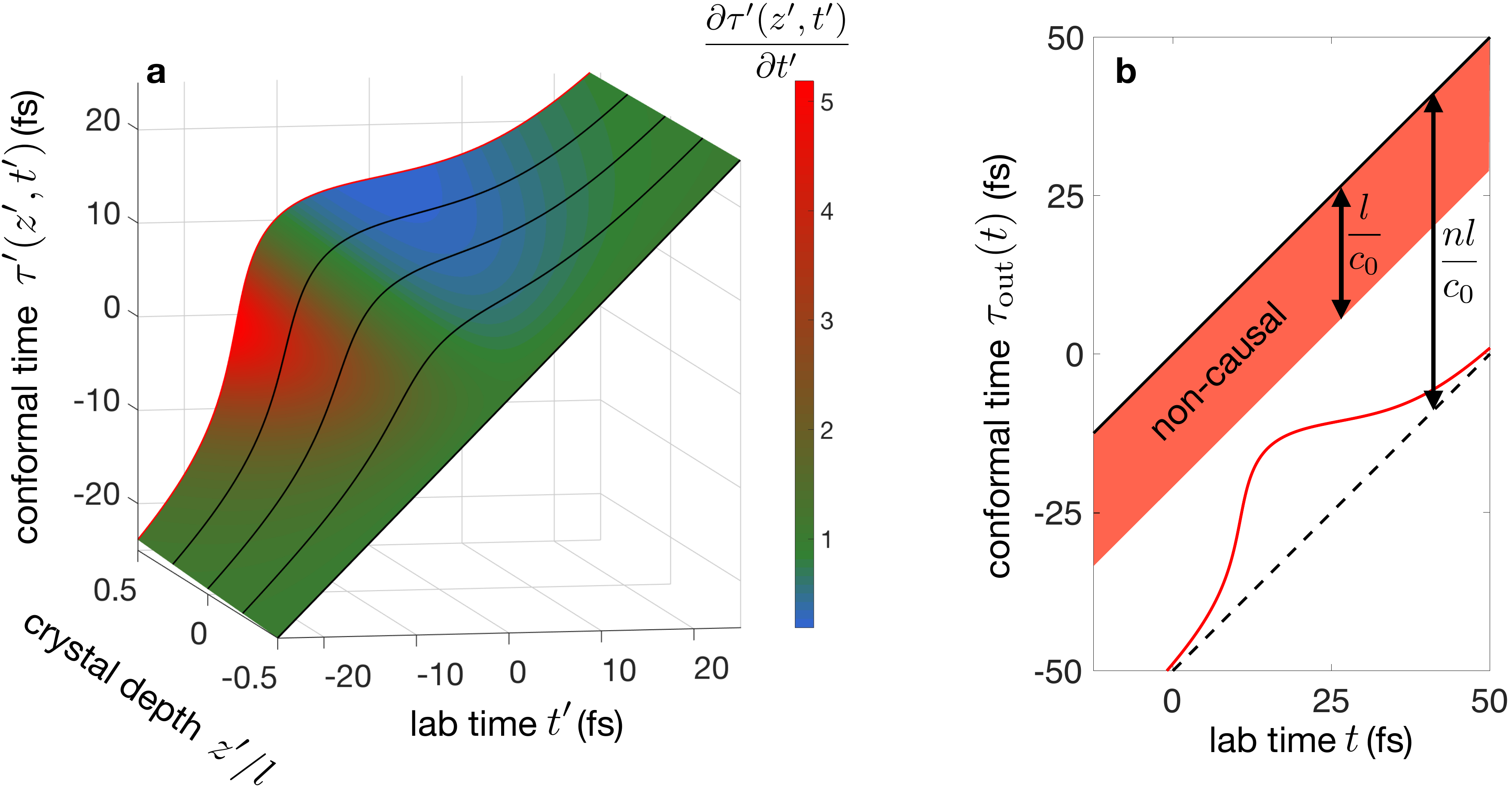}
\caption{ {\bf Behaviour of the conformal time with respect to the lab time illustrated for the half-cycle pulse [cf. Eq.~\eqref{HCP}]
 with $r=5$ and $\Gamma_0/(2\pi)=26$ THz.  a}, Conformal time $\tau'(z',t')$ as a function of the lab time $t'$ and the propagation length inside the crystal $z'=z$. The retarded reference frame is used. $\tau'(z',t')$ coincides with $t'$ at the entrance of the crystal $z'=-l/2$ and starts to deviate from it for $z'>-l/2$. The graph is coloured according to the values of the inverse conformal factor $\frac{\partial\tau'(z',t')}{\partial t'}$. In the simplified picture, which does not yet incorporate the detection process influenced also by the change in the local density of the world lines, departure from green towards blue (red) leads to squeezing (anti-squeezing) of the vacuum fluctuations.  The black lines help the visualization of the surface. \textbf{b}, Red line: final conformal time  $\tau_{\mathrm{out}}(t)=\tau(z=l/2,t)$ at the exit of the crystal as a function of the lab time $t$, shown in the original reference frame. Dashed black line: same without the driving field  --- delayed by $nl/c_0$ with respect to $t$. Full black line: the run of the lab time $t$ is shown for comparison. Values of $\tau_{\mathrm{out}}(t)$ must stay below the red area  defined by the delay time $t_{\mathrm{d,0}}=l/c_0$ in order not to violate causality (cf. Supplementary Information).\label{Fig1}}
\end{figure}

We consider the setup depicted schematically in Fig.~\ref{Fig0}a. A coherent mid-infrared electric driving field with a classical amplitude $\mathcal{E}(z,t)$ propagates along the $z$-axis and is polarized along the $x$-axis. It enters a thin transparent optical crystal, which has a length $l$, centered at $z=0$, and possesses a non-vanishing second-order nonlinearity. This leads to an interaction of the driving field with the co-propagating quantum electric field component $\hat{\varepsilon}(z,t)$, which is polarized along the $y$-axis, belongs to the same frequency range, and corresponds to the vacuum field when entering the crystal (see Methods for
details). The nonlinear interaction in the crystal subsequently modifies $\hat{\varepsilon}(z,t)$.  Far from material resonances the second-order susceptibility $\chi^{(2)}$ can be treated as a dispersionless tensor. Adopting the propagation of plane waves with perfect phase matching inside the crystal and a broadband version of the slowly varying amplitude approximation the evolution of the field operator $\hat{\varepsilon}(z,t)$ is given by (see Methods)
\begin{align}
\frac{\partial \hat{\varepsilon}}{\partial z}&=\frac{d}{nc_0}\left[\frac{\partial \mathcal{E}}{\partial t} \hat{\varepsilon}+\left(\mathcal{E}-\frac{n^2}{d}\right)\frac{\partial\hat{\varepsilon}}{\partial t}\right].\label{PDE}
\end{align}
Here the effective nonlinear coefficient $d$ is determined by the properties of $\chi^{(2)}$, $n$ is the linear refractive index and $c_0$ is the vacuum speed of light. Equation \eqref{PDE} can be solved analytically by using the method of characteristics, where the initial conditions of a partial differential equation are propagated along   \textit{characteristic curves}  (see Methods for
details). In our case the characteristic curves are given by fixed values of an auxiliary time variable $\tau(z,t)$, which at any fixed $z$ monotonically increases with $t$ and may be called \textit{conformal time,} in analogy to cosmology~\cite{Mukhanov_book}. The gravitational analogues of the characteristic curves are world lines \cite{Guedes2018}. Denoting the quantum field at the entrance of the generation crystal as $\hat{\varepsilon}_\mathrm{in}(t)\equiv\hat{\varepsilon}(z=-l/2,t)$, the solution of equation \eqref{PDE} can be found as
\begin{align}
\hat{\varepsilon}(z,t)=\frac{\partial\tau(z,t)}{\partial t}\hat{\varepsilon}_\mathrm{in}(\tau(z,t)).\label{Sol}
\end{align}
The form of the resulting quantum field
can be attributed to two different effects. Firstly, due to the driving field $\mathcal{E}(z,t)$ the wave propagation of the incoming quantum field $\hat{\varepsilon}_\mathrm{in}$ inside the crystal is governed by the conformal time $\tau(z,t)$. This effect is illustrated by the world lines in the horizontal plane of Fig.~\ref{Fig0}b.
The change of their density upon the propagation through the crystal determines the resulting difference in the flow of the conformal time with respect to the lab time $t$. Secondly, there is a modulation of the amplitude of the quantum field by the inverse conformal factor \cite{Mukhanov_book} $\frac{\partial\tau(z,t)}{\partial t}$. This factor is shown in the vertical plane of Fig.~\ref{Fig0}b at the crystal exit and is essentially related to the behaviour of $\frac{\partial \mathcal{E}(z,t)}{\partial t}$ (cf. ref. \onlinecite{Riek2017}), as discussed below. Note that the quantum field is enhanced (suppressed) in the same time segments where the density of the world lines is increased (decreased), see vertical lines in Fig.~\ref{Fig0}b. In a simplified picture, neglecting the effect of the modified density of the world lines, the amplitude modulation of the quantum field leads to a corresponding modulation in the temporal profile of its quantum noise. The quantum fluctuations of $\hat{\varepsilon}$ are suppressed beneath the vacuum level in certain time segments while exceeding it in the neighbouring segments. Squeezing (anti-squeezing) can be connected to the deceleration (acceleration) of the local flow of the conformal time $\tau(z,t)$, determined by its derivative with respect to the lab time $t$. For a part of the discussion it is convenient to use a retarded reference frame with $t'=t-\frac{n}{c_0}z$, $z'=z$ and $\tau'(z',t')=\tau(z,t)+\frac{nl}{2c_0}$. First we discuss the essence of the effect and its detection for the case of an idealized half-cycle driving pulse; afterwards we study the single-cycle case. Figure \ref{Fig1}a depicts the evolution of the conformal time in the retarded reference frame through the crystal resulting from a half-cycle driving field of the form (see Supplementary Information)
\begin{align}
 \mathcal{E}'(t')&=\mathcal{E}_0\sech(\Gamma_0t'), \label{HCP}
\end{align}
where $\mathcal{E}_0$ is
the amplitude of the field and $\Gamma_0$ determines its duration.
The squeezing strength $r=|\mathcal{E}_0d|\Gamma_0 l/(nc_0)$ is a dimensionless positive parameter. At the entrance of the crystal the conformal time matches the lab time: $\tau(z=-l/2,t)=t$ and $\tau'(z'=-l/2,t')=t'$. With the interaction turned on, while the quantum field propagates through the crystal its conformal time starts to deviate from the lab time. 

 Via Eq.~\eqref{Sol} the conformal time directly links the quantum field at the exit of the crystal at time $t$ to the quantum field at the entrance at time $\tau_{\mathrm{out}}(t)=\tau(z=l/2,t)$, see Fig. \ref{Fig1}b. According to Eq.~\eqref{Sol} the modification of the variance of the quantum electric field after passing the crystal is given by  $\left(\mathrm{d} \tau_{\mathrm{out}}/\mathrm{d} t\right)^2$ if considered at a fixed time moment. Experimentally,  the temporal dynamics of the quantum fluctuations of the electric field in dependence on the relative time delay of the probe pulse $t_\mathrm{d}$ can be traced utilizing the electro-optic sampling technique and statistical readout \cite{Riek2015,Riek2017}. The time resolution and sensitivity are limited by the duration of the probe pulse $t_\mathrm{p}$ during which the information about the quantum field is collected. This is a significant aspect because apart from the change in its amplitude the quantum field of the incoming vacuum is effectively subjected to a modified flow of time at the crystal exit, as can be seen from Eq.~\eqref{Sol}. Alternatively we can say that in the reference frame using the conformal time $\tau_{\mathrm{out}}(t)$ in place of the lab time $t$ the probe pulse shape and duration are modified according to the flow of $\tau_{\mathrm{out}}(t)$ [cf. Eq.~\eqref{Eq:inverse_time_flow} in Methods]. In order to obtain the time-resolved detected variance $V(t_\mathrm{d})=\left\langle[\hat{\varepsilon}^{\mathrm{(d)}}_{\mathrm{out}}(t_\mathrm{d})]^2\right\rangle$ for the outgoing quantum electric field $\hat{\varepsilon}_\mathrm{out}(t)=\hat{\varepsilon}(z=l/2,t)$, i.e. the temporal trace of its quantum fluctuations, we employ the quantum theory of electro-optic sampling \cite{Moskalenko2015} (see Methods for details). We define the relative detected variance (RDV) as
\begin{align}
\mathrm{RDV}(t_\mathrm{d})&=\frac{V(t_\mathrm{d})
-(\Delta \hat{\varepsilon}_\mathrm{vac})^2}{(\Delta \hat{\varepsilon}_\mathrm{vac})^2}.\label{Def}
\end{align}
Here $(\Delta \hat{\varepsilon}_\mathrm{vac})^2=\left\langle[\hat{\varepsilon}^{\mathrm{(d)}}_{\mathrm{in}}]^2\right\rangle$ denotes the detected variance of the incoming quantum vacuum field, which does not depend on the delay time $t_\mathrm{d}$.

\begin{figure}[t!]
\centering
\includegraphics[scale=0.4]{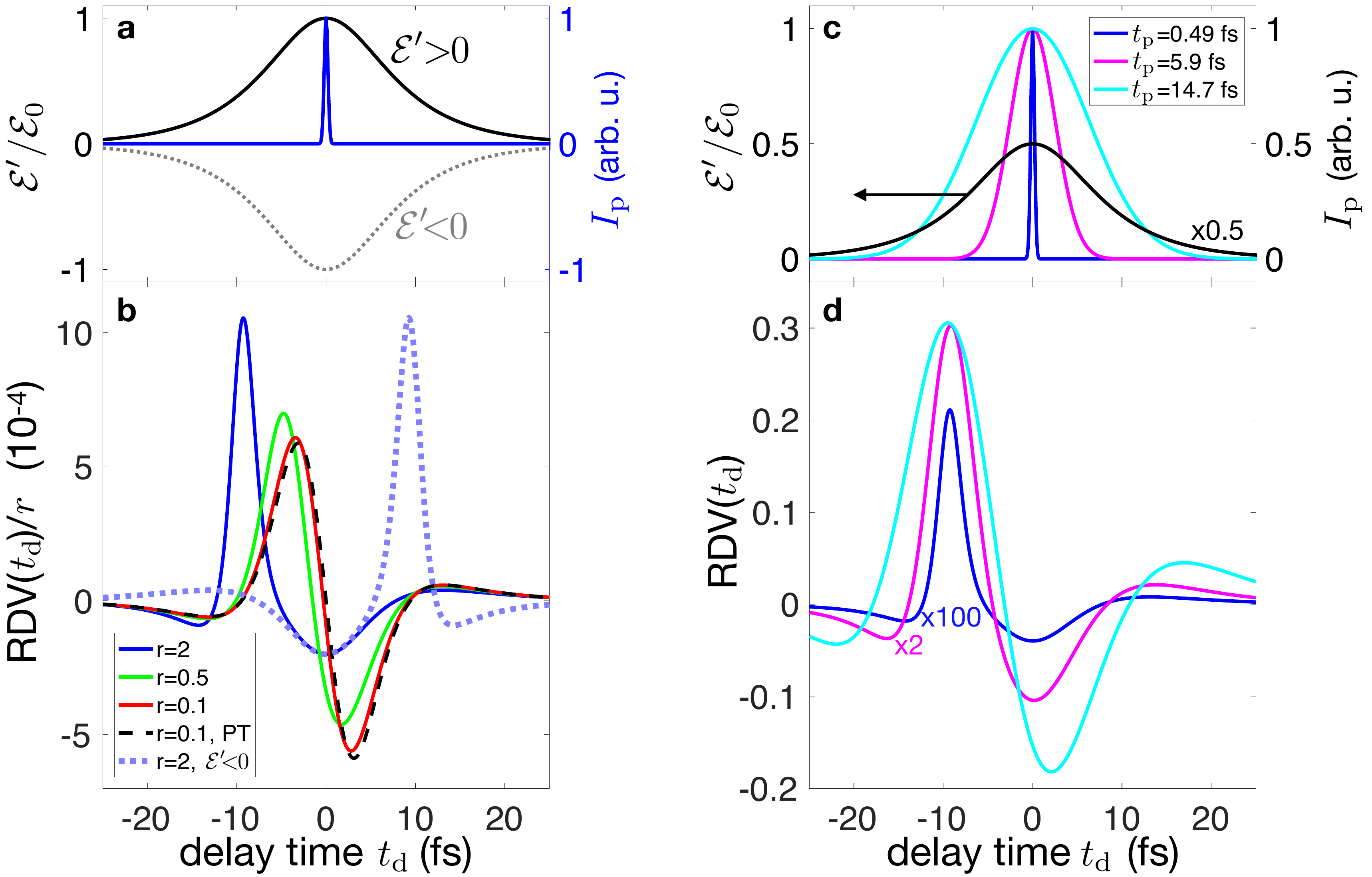}
\caption{
{\bf Relative detected variance (RDV) in dependence on the strength of the half-cycle driving field and probe pulse duration. a}, Temporal profiles of the driving field $\mathcal{E}'$ with two opposite polarities and  probe pulse intensity envelope $I_\mathrm{p}$ with duration $t_\mathrm{p}=0.49$ fs (blue). \textbf{b}, Dynamics of the normalized RDV for a fixed probe pulse duration $t_\mathrm{p}=0.49$~fs and different squeezing strengths $r$=0.1 (red), 0.5 (green) and 2 (blue), proportional to the driving field amplitude. The normalization by $r$ is chosen to keep the signal magnitude in the same range. For comparison, the exact analytical result within the first-order perturbation theory (PT) in $r$ and limit of vanishing $t_\mathrm{p}$ is shown (black dashed line). The light blue dotted line shows the RDV for the half-cycle pulse with $\mathcal{E}'\!<\!0$ for $r=2$. \textbf{c}, Temporal profiles of the driving field $\mathcal{E}'$ (black, normalized by its amplitude $\mathcal{E}_0$) and probe pulse intensity envelope $I_\mathrm{p}$ for different probe pulse durations $t_\mathrm{p}=0.49$~fs (blue), $5.9$ fs (magenta) and $14.7$ fs (cyan) are shown on the same time scale as the RDV.
\textbf{d}, Dynamics of the RDV for a fixed $r=2$ and the same probe pulse durations as in the upper panel.
 \label{Fig2}}
\end{figure}
Figure \ref{Fig2}b depicts the RDV for the case of the half-cycle pulse Eq.~\eqref{HCP}, a very short detection time $t_\mathrm{p}=0.49$~fs and different squeezing strengths $r$.  
For $r=0.1$,  RDV($t_\mathrm{d}$)
is an almost perfectly odd function (red line) with respect to the center of the driving pulse at $t_\mathrm{d}=0$~fs. In this case, the temporal trace nearly coincides  with the waveform proportional to the third derivative of the driving field $\mathrm{d}^3 \mathcal{E}'(t_\mathrm{d})/\mathrm{d} t_\mathrm{d}^3$ (dashed black line). The latter corresponds to the exact analytic solution in the limit of small $r$ and $t_\mathrm{p}$, which can be obtained via the squeezing operator in the frequency domain (see Supplementary Information). This dynamics differs from the simplified picture where for each fixed time moment the variance of the quantum field is determined by  $\left[\mathrm{d} \tau_{\mathrm{out}}/\mathrm{d} t\right]^2$ and in the low squeezing regime ($r\ll1$) leads to temporal profiles following the first derivative $\mathrm{d} \mathcal{E}'(t_\mathrm{d})/\mathrm{d} t_\mathrm{d}$ \cite{Riek2017}. Notice that the deviation becomes less significant if multi-cycle, more narrowband driving fields are considered.

For small $r$, the RDV scales linearly with $r$ and therefore remains symmetric. A build-up of asymmetry between the time segments with reduced and excess quantum noise can be clearly observed when $r$ is increased. Firstly, we can see that the magnitude of the detected quantum fluctuations becomes more pronounced in the anti-squeezing period with respect to the squeezing period. This can be attributed to the fact that the conformal time $\tau_\mathrm{out}(t)$ must always increase monotonically with $t$. As the slope of $\tau_\mathrm{out}(t)$ approaches zero, the flow of time comes to a halt and according to  Eq.~\eqref{Sol} $\hat{\varepsilon}_\mathrm{out}(t)$ must vanish. In the simplified picture, looking only at the prefactor in Eq.~\eqref{Sol}, this would lead to an almost complete elimination of the quantum noise and squeezing limited by 100\%. In contrast, acceleration in the flow of time in principle may lead to arbitrarily large prefactors determining the magnitude of $\hat{\varepsilon}$ in Eq.~\eqref{Sol}  and therefore to an arbitrarily strong enhancement of the quantum noise. In the full picture including detection, the increase caused by the prefactor in Eq.~\eqref{Sol} is partly counteracted by the slower (quicker) flow of  the conformal time  in $\hat{\varepsilon}_\mathrm{in}(\tau_\mathrm{out}(t))$ in the case of squeezing (anti-squeezing). The duration of the probe pulse seen by the incoming vacuum field thus effectively  becomes smaller (larger) for squeezing (anti-squeezing) and the detected variance $V(t_\mathrm{d})$ increases (decreases) due to this effect \cite{Riek2015}. Equivalently, the effect of the lower (higher) local world line densities (see Fig.~\ref{Fig0}b) alone would enhance (suppress) the detected quantum noise.  Still, after the unperturbed vacuum contribution $(\Delta \hat{\varepsilon}_\mathrm{vac})^2$ is subtracted in Eq.~\eqref{Sol}, the asymmetry in the RDV($t_\mathrm{d}$) is preserved to a large extent for sufficiently short detection times $t_\mathrm{p}$. Secondly, with increasing $r$ the time segments of squeezing become broader while the time segments of anti-squeezing narrow down.
This additional asymmetry can be comprehended looking at Fig.~\ref{Fig1}b. It is clear that the time intervals with steep slopes in the flow of the conformal time $\tau_\mathrm{out}(t)$ (anti-squeezing) take less space on the horizontal axis, representing the lab time $t$, than the intervals with flat slopes (squeezing). This is what we observe for the case of r=2 in Fig. 3b (blue line). Uniquely to the strong squeezing in the time domain, polarity reversal of the driving field (Fig. 3a) does not change the position of the squeezing valley while reverses the arrival time of the anti-squeezing burst.

In Fig.~\ref{Fig2}d, the dynamics of the RDV is shown for different probe pulse durations $t_\mathrm{p}$ and a fixed squeezing strength $r=2$. As $t_\mathrm{p}$ grows (cf. Fig. \ref{Fig2}c), higher frequencies in $\hat{\varepsilon}_\mathrm{out}(t)$ are not captured anymore leading to a flattening of the detected temporal traces. The narrow anti-squeezing peaks are affected more strongly by such changes than the valley regions of squeezing. When the probe pulse duration approaches the oscillation time of the driving field, the asymmetry is almost completely lost.

\begin{figure}[t]
\centering
\includegraphics[scale=0.4]{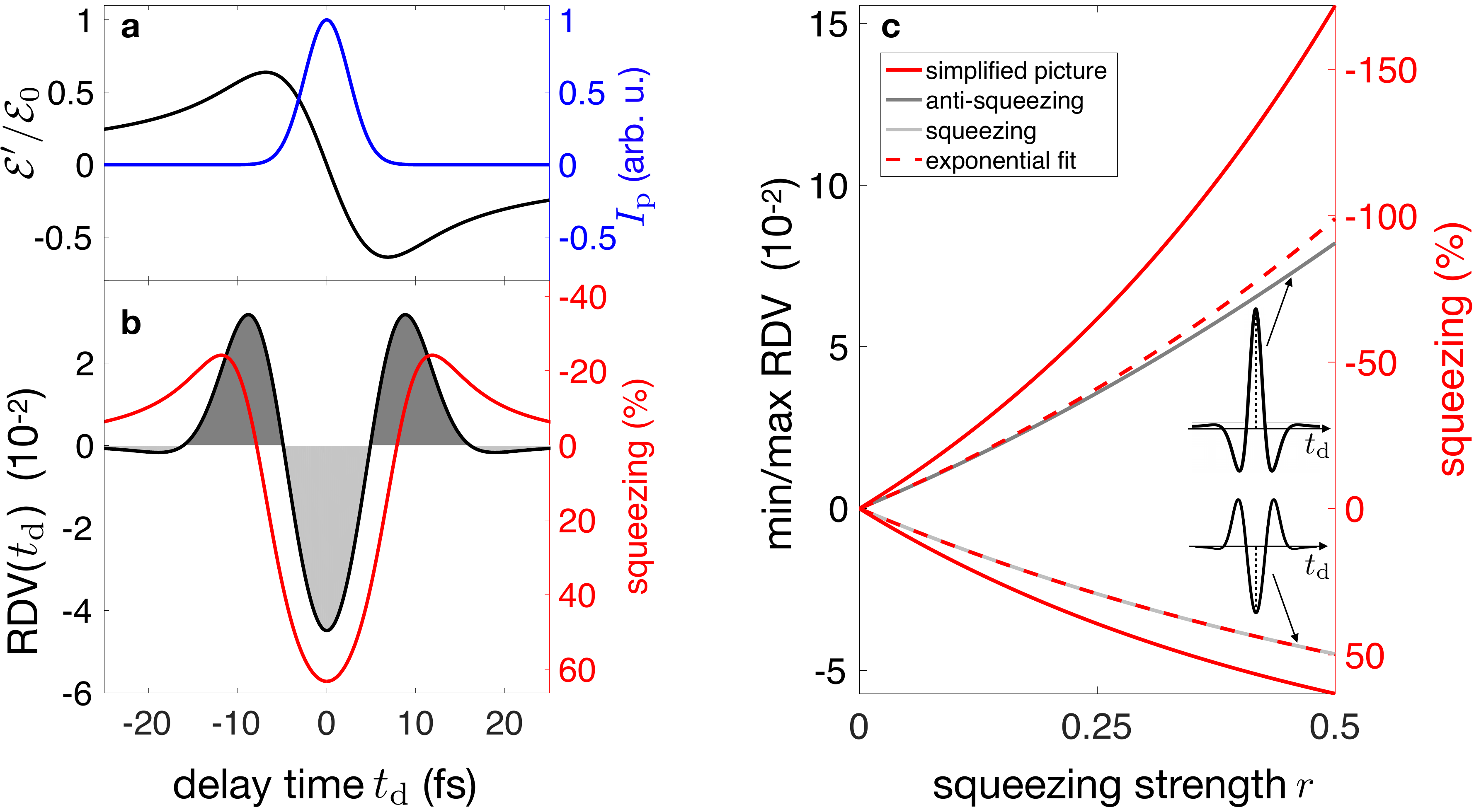}
\caption{ {\bf Pulsed squeezing for single-cycle driving.
a}, Temporal profiles of the driving field $\mathcal{E}'(t')=\mathcal{E}_0\left[\exp\left(-\Gamma_0^2 t'^2\right)-1\right]/(\Gamma_0t')$ with $\Gamma_0/(2\pi)= 26$ THz (black) and probe pulse intensity envelope $I_\mathrm{p}$ with  $t_\mathrm{p}=5.9$~fs (blue).
\textbf{b}, Corresponding dynamics of the RDV for $r=0.5$ (black line).
Dark grey (light grey) areas denote anti-squeezing (squeezing).
The red line depicts the noise trace obtained within the simplified picture, where the degree of squeezing (right axis) can be extracted directly from the ratio between the variances of the outgoing and incoming quantum field at each $t_\mathrm{d}=t$. Negative values for the degree of squeezing correspond to anti-squeezing. \textbf{c},
Values of the RDV at $t_\mathrm{d}=0$~fs plotted against $r$ for two different polarities of the driving field. Depending on the polarity, it is the maximum (anti-squeezing, dark grey line, upper inset) or the minimum (squeezing, light grey line, lower inset) value of the RDV.
The  degree of squeezing (dashed red lines) results from an exponential fit (cf. Supplementary Information). The RDV in \textbf{b} is then rescaled according to this fit to obtain the degree of squeezing at arbitrary times.
The red lines in \textbf{c} show the degree of squeezing calculated within the simplified picture. For a vanishing probe pulse duration $t_\mathrm{p}\rightarrow 0$, the squeezing and anti-squeezing curves obtained from the RDV converge towards this result.
 \label{Fig3}}
\end{figure}

We now switch our attention to the case of a single-cycle driving pulse (Fig. \ref{Fig3}a).
Figure \ref{Fig3}b shows the resulting dynamics of the RDV, also in terms of the degree of squeezing (see Supplementary Information). For the given amplitude of the driving field, the maximum positive degree of squeezing amounts to $49.6\%$ while the minimum negative degree of squeezing (corresponding to anti-squeezing) constitutes only $-35\%$. Therefore, the previously described asymmetry in favour of higher absolute values for anti-squeezing is reversed in this case. At first sight, one might wonder whether Heisenberg's uncertainty principle is violated in the described situation. Through the inspection of the RDV trace for the polarity-switched drive (inset in Fig. \ref{Fig3}c), we discover that the amplitude of anti-squeezing is in fact nearly twice that of the squeezing (Fig. \ref{Fig3}b). It is only in this temporally non-local sense that we are able to restore the familiar dependence of quantum noise on the squeezing strength (Fig. \ref{Fig3}c and Supplementary Information), and hence directly recover Heisenberg's uncertainty  principle.

Our work provides a consistent time-domain theory of the generation and sub-cycle-resolved detection of ultrabroadband waveforms of pulsed squeezed radiation, linking these processes directly to a change in the local flow of time induced by the coherent driving field. This constitutes a non-perturbative analytical solution for the operator of the generated quantum field and the electro-optically sampled traces of its quantum noise. This solution is valid at any time and also in the high-driving regime, restricted only by conditions underlying the broadband version of the slowly varying amplitude approximation and causality. We applied our theory to predict time traces of the detected variance and corresponding degrees of squeezing belonging to ultrashort quantum fields created by half-cycle and single-cycle driving.
The results for the detected variance show that the use of electro-optic sampling for a time-resolved measurement of the noise pattern of a squeezed field inherently introduces an admixture of vacuum fluctuations. However, the asymmetries between squeezing and anti-squeezing are preserved by this measurement technique for sufficiently short probe pulses.
 Finally, we theoretically predict an effect that at first glance looks paradoxical: the conventionally observed asymmetry between squeezing and anti-squeezing can be reversed for specially designed driving fields.

\newpage
\section*{Methods}
\subsection*{Time-domain equation for the generation process}
We consider a process where in a $\chi^{(2)}$ nonlinear crystal  a coherent driving field $\mathcal{E}(z,t)$ polarized along the $x$-axis propagates in the $z$-direction (cf. Fig.~\ref{Fig0}a). $\mathcal{E}(z,t)$ interacts with the $y$-component of a co-propagating quantum field $\hat{\epsilon}(z,t)$. This gives rise to the $y$-component of the nonlinear polarization
 $\hat{P}^{(2)}(z,t)=-\epsilon_0 d\mathcal{E}(z,t)\hat{\epsilon}(z,t)$
but generates no $x$-component. Such a situation can be realized for various nonlinear crystals, e.g., for ZnTe in a configuration where the present unit vectors $\textbf{e}_x$ and $\textbf{e}_z$  would be aligned along the [001] and [110] crystallographic directions, respectively \cite{Planken2001}. The driving field can be external but the situation is similar if this field is generated in the same crystal \cite{Riek2017}.
We write the corresponding inhomogeneous wave equation for $\hat{\epsilon}(z,t)$ with $\hat{P}^{(2)}(z,t)$ as the source term. Then we decompose $\hat{\epsilon}(z,t)=\int_{-\infty}^{\infty} \mathrm{d}\Omega\; \hat{\varepsilon}(z;\Omega)e^{i(k_\Omega z-\Omega t)}$ and in the same way $\mathcal{E}(z,t)$ and $\hat{P}^{(2)}(z,t)$ in plane waves with an almost ideally linear dispersion $k_\Omega=\frac{n}{c_0}\Omega$.
 With the latter, for a thin enough crystal perfect phase matching is a good approximation.
In the resulting frequency-domain integro-differential equation \cite{Riek2017}
we make an assumption that
the term $\frac{\partial^2 \hat{\varepsilon}(z;\Omega)}{\partial z^2} $ can be neglected in comparison with $ k_\Omega \frac{\partial \hat{\varepsilon}(z;\Omega)}{\partial z}$. This leads to
\begin{equation}
\frac{\partial \hat{\varepsilon}(z;\Omega)}{\partial z}=-\frac{id\Omega}{n c_0} \int^{\infty}_{-\infty}d\Omega_1 \mathcal{E}^*(z;\Omega_1-\Omega)\hat{\varepsilon}(z;\Omega_1).
\label{field_eq}
\end{equation}
Transforming Eq.~\eqref{field_eq} back into the time-domain we then obtain Eq.~\eqref{PDE}.

 In essence, this step is similar to the well-known slowly varying amplitude approximation  (SVAA) \cite{Boyd_book}. Note, however, that in our case no amplitude or envelope was introduced in the time domain. Therefore, also few-cycle and subcycle fields with no unambiguously defined carrier frequency and phase can be described by this approach. It  may be considered as a broadband version of the SVAA\cite{Loudon_book}. An important prerequisite is the assumed linear dispersion holding in the frequency range of interest.
In a dispersive medium ultrashort pulses would experience shape distortion while propagating through the crystal \cite{Brabec1997}, requiring a more involved description of the generation process.

\subsection*{Method of characteristics}
The method of characteristics is used to solve a given partial differential equation (PDE) by reducing it to a set of ordinary differential equations (ODEs), which define parametric curves  in the space of the PDE variables (so-called characteristic curves). These curves allow then to construct the solution of the PDE.
Let us assume that the original variables of the PDE, in our case Eq.~\eqref{PDE}, depend on the parameter $\zeta$. Comparison of the total derivative of $\hat{\epsilon}$ with respect to $\zeta$ with Eq.~\eqref{PDE} gives us ODEs for $z(\zeta)$ and $t(\zeta)$ (with their respective initial conditions):
\begin{align}
&\frac{\mathrm{d}z}{\mathrm{d}\zeta}=1,  \qquad z(\zeta=0)=-l/2;\label{ODEz}\\
&\frac{\mathrm{d}t}{\mathrm{d}\zeta}=-\frac{d}{nc}\left[\mathcal{E}(z,t)-\frac{n^2}{d}\right]_{z=z(\zeta),t=t(\zeta)},
\qquad t(\zeta=0)=\tau. \label{ODEt}
\end{align}
Here an additional parameter $\tau$ was introduced as the initial condition for $t(\zeta)$ when $\zeta=0$. Solving Eq.~\eqref{ODEz} reveals that $\zeta=z+l/2$. Changing the variable from $\zeta$ to $z$ in Eq.~\eqref{ODEt} and solving it, we obtain $t(z,\tau)$, where now also an explicit dependence on $\tau$ appears. Finally, the function $t(z,\tau)$ can be inverted to obtain $\tau(z,t)$ with $\tau(z=-l/2,t)=t$.
The functional form of $\tau(z,t)$ ensures that the total derivative of $\tau(z(\zeta),t(\zeta))$ with respect to $\zeta$ vanishes,
\begin{align*}
\frac{\mathrm{d} \tau(z(\zeta),t(\zeta))}{\mathrm{d}\zeta}&=\frac{\partial \tau(z,t)}{\partial z}-\frac{d}{nc_0}\left[\mathcal{E}(z,t)-\frac{n^2}{d}\right]\frac{\partial \tau(z,t)}{\partial t}=0,
\end{align*}
meaning that $\tau(z(\zeta),t(\zeta))$ remains constant when moving along a characteristic curve $\{z(\zeta),t(\zeta)\}$ parameterized by $\zeta$.
This PDE for $\tau(z,t)$ can now be used to prove that Eq.~\eqref{Sol} indeed satisfies Eq.~\eqref{PDE}. Therefore, the only remaining nontrivial task is to solve Eq.~\eqref{ODEt} finding $\tau(z,t)$. Inserting the solution for $\tau(z,t)$ into Eq.~\eqref{Sol} then gives $\hat{\varepsilon}(z,t)$.

 For the purpose of solving Eq.~\eqref{ODEt} the retarded reference frame with $t'=t-\frac{n}{c_0}z$, $z'=z$ and $\mathcal{E}(z,t)=\mathcal{E}'(t')$ is utilized. This transforms Eq.~\eqref{ODEt} into a separable ODE. $\tau(z,t)$ can then be expressed as
\begin{align*}
\tau(z,t)&=f^{-1}\left(f\Big(t-\frac{n}{c_0}z\Big)+\frac{1}{nc_0}\Big(z+\frac{l}{2}\Big)\right)-\frac{nl}{2c_0}
\end{align*}
in the original reference frame, where $f(t')$ is defined as solution of the following ODE
\begin{align*}
\frac{\mathrm{d}f(t')}{\mathrm{d}t'}=\frac{1}{d\mathcal{E}'(t')}.
\end{align*}
Note that here no initial condition needs to be stated because it can be chosen arbitrarily. The necessary initial condition is already incorporated in the solution for $\tau(z,t)$ with $\tau(z=-l/2,t)=t$, regardless of the specific form of $f(t')$.

\subsection*{Quantum electro-optic sampling}
The quantum theory of electro-optic sampling \cite{Moskalenko2015} is employed in order to model the detection process. The variance of the outgoing quantum field $\hat{\varepsilon}'_\mathrm{out}(t')$ is calculated as a function of the relative delay time $t_\mathrm{d}$ with respect to the center of an ultrashort higher-frequency probe pulse $\mathcal{E}'_\mathsf{p}(t')$ (for convenience the retarded reference frame is used here). The resulting operator of the sampled field is given by
\begin{align}
\hat{\varepsilon}^{\mathrm{(d)}}_{\mathrm{out}}(t_\mathrm{d})=\int_{-\infty}^\infty \!\mathrm{d}t'\; R(t_\mathrm{d}-t')\hat{\varepsilon}'_\mathrm{out}(t').\label{Det}
\end{align}
The detector function $R(t_\mathrm{d})$ is determined by the normalized intensity of the probe field
that for a fast oscillating probe field  can be replaced by the normalised intensity envelope:
\begin{align*}
R(t_\mathrm{d})=\frac{\left|\mathcal{E}'_\mathsf{p}(t_\mathrm{d})\right|^2}{\int_{-\infty}^\infty \!\mathrm{d}t'\, \left|\mathcal{E}'_\mathsf{p}(t')\right|^2}\approx\frac{I'_\mathsf{p}(t_\mathrm{d})}{\int_{-\infty}^\infty\!\mathrm{d}t'\, I'_\mathsf{p}(t')}.
\end{align*}
Then for the outgoing quantum field, determined by Eq.~\eqref{Sol} at $z=l/2$, the operator of the sampled field can be expressed via the incoming quantum field $\hat{\varepsilon}_\mathrm{in}(t)$ in the original reference frame as
\begin{align}\label{Eq:inverse_time_flow}
\hat{\varepsilon}^{\mathrm{(d)}}_\mathrm{out}(t_\mathrm{d})=\int_{-\infty}^\infty \mathrm{d}t\, R\big(t_\mathrm{d}-\tau^{-1}_\mathrm{out}(t)\big)\hat{\varepsilon}_\mathrm{in}(t),
\end{align}
where $x=\tau^{-1}_\mathrm{out}(y)$ denotes the inverse function of $y=\tau_\mathrm{out}(x)$.
Let us consider the case where the incoming field corresponds to the vacuum state.
Then Eq.~\eqref{Eq:inverse_time_flow} shows that sampling of the generated squeezed field by the probe pulse in an inertial reference frame
can be alternatively viewed as sampling of the bare vacuum field in a reference frame with a non-inertial time flow given by $\tau^{-1}_\mathrm{out}(t)$. In the latter reference frame the shape of the used probe pulse is then effectively transformed.
In order to characterize the dynamics of the resulting quantum fluctuations we evaluate the relative detected variance (RDV) that is given by Eq.~\eqref{Def}. For the corresponding calculations in this paper we used
 a Gaussian shape of the probe intensity envelope with the FWHM duration $t_\mathrm{p}$ so that $R(t_\mathrm{d})=\frac{2\sqrt{\ln2}}{\sqrt{\pi}t_\mathrm{p}}\exp\left(-\frac{4\ln2\,t_\mathrm{d}^2}{t_\mathrm{p}^2}\right)$.  
 
The calculated values of the RDV typically are small compared to the values of RDV and respective degrees of squeezing that would follow directly from the simplified picture (see Fig. \ref{Fig3}b).
The reason for this is that the RDV [cf. Eq.~\eqref{Def}] is defined as the relative difference between the detected variance of the generated field and that of the unperturbed vacuum fluctuations.
In order to resolve the signal in the time-domain the probe duration has to be much shorter than the typical time scales of the signal. Thus the frequency spectrum of the probe has to be much broader than that of the coherent field driving the squeezing. This means that the probe pulse will not only sample modes at the frequencies that are affected by the driving field but also modes at higher frequencies which remain untouched. Therefore, the contribution of the unperturbed vacuum fluctuations to the RDV has to be significantly larger than the contribution from the frequency range affected by the generation process in order to appropriately resolve the temporal squeezing pattern.
 This can be interpreted as an admixture of vacuum fluctuations to the signal that can be modelled by losses. However, it is important to note that rather than being losses that can be avoided, these losses are inherent to the measurement process and are necessary in order to appropriately resolve the signal in time. Nevertheless,  losses do not necessarily represent an obstacle for the reconstruction of the degree of the generated squeezing \cite{Riek2017}, which is shown in Fig. \ref{Fig3}c.

Note that the relevant quantity measured experimentally, and connected to the calculated RDV, is the relative differential noise (RDN) \cite{Riek2017}. It is given by
\begin{align*}
\mathrm{RDN}(t_\mathrm{d})&=\left(\sqrt{V(t_\mathrm{d})}-\Delta\hat{\varepsilon}_\mathrm{vac}\right)\frac{\Delta\hat{\varepsilon}_\mathrm{vac}}{\left(\Delta\mathcal{E}_\mathrm{p,SN}\right)^2}\;,
\end{align*}
[see Eq.~\eqref{Def} for comparison].
Here $\left(\Delta\mathcal{E}_\mathrm{p,SN}\right)^2$ denotes the shot noise (SN) of the probe field $\mathcal{E}_\mathrm{p}$.

\section*{Acknowledgements}
We thank P. Sulzer and R. Haussmann for discussions. Support
by DFG via SFB767, by Baden-Württemberg Stiftung via
the Eliteprogramme for Postdocs
(project ``Fundamental aspects of relativity and causality in time-resolved quantum optics '')
and by Young Scholar Fund of the University of Konstanz is acknowledged.
M.K. is indebted to the LGFG PhD fellowship program of the University of Konstanz.

\section*{Author Contributions}
A.S.M., D.V.S. and G.B. conceived the idea. A.S.M. managed the project and supervised the research. M.K. found the exact analytical solution in the time domain, performed numerical calculations and prepared the figures. T.L.M.G. obtained the perturbative analytic solution via the squeezing operator in the frequency domain. M.K., T.L.M.G. and A.S.M.  wrote the first version of the paper.  D.V.S. and A.L. provided several important physical insights and interpretations. All authors discussed the results and contributed to the writing of the final manuscript.

\renewcommand{\thefigure}{S\arabic{figure}}
\setcounter{figure}{0}

\renewcommand{\theequation}{S\arabic{equation}}
\setcounter{equation}{0}

\section*{Supplementary information}

\subsection*{Slowly varying amplitude approximation and causality}
Equation~\eqref{Sol} shows that the quantum field at the exit of the crystal at time $t$ is influenced by the quantum field at the entrance at time $\tau_\mathrm{out}(t)$. This fact raises the question whether in our calculations this can technically lead to an information exchange faster than
 the speed of light or even the conformal time preceding the lab time. Already the first case would contradict causality. Figure \ref{Fig1}b shows $\tau_{\mathrm{out}}(t)$  as a function of $t$ (full black line). Without the driving field, $\tau_\mathrm{out}(t)$ is delayed with respect to $t$ by the time $nl/c_0$ which the light needs to travel through the crystal (see dashed black line). The red-coloured area in the figure marks the region for $\tau_{\mathrm{out}}(t)$ where the delay between $\tau_{\mathrm{out}}(t)$ and $t$ is less than $l/c_0$, which is the time required for light to travel the distance $l$ in vacuum. Due to causality, the information about the incoming quantum field cannot propagate to the end of the crystal in a time shorter than $l/c_0$. Causality forbids this area for $\tau_{\mathrm{out}}(t)$. Since its deviation from the lab time $t$ grows with higher squeezing strengths $r$, the restriction that $\tau_{\mathrm{out}}(t)-t+nl/c_0$ (vertical difference between the full red and dashed black line in Fig.~\ref{Fig1}b) should not exceed $(n-1)l/c_0$ ultimately puts an upper bound to the maximally allowed $r$. However, the use of a broadband version of the SVAA, underlying our theoretical approach, imposes an even stricter upper bound to the squeezing strength that then already ensures causality, as we will show in the following. 
Within the utilized broadband SVAA we neglected $\frac{\partial^2 \hat{\varepsilon}(z;\Omega)}{\partial z^2} $ with respect to $ k_\Omega \frac{\partial \hat{\varepsilon}(z;\Omega)}{\partial z}$. This is possible only if $\hat{\varepsilon}(z;\Omega)$ does not change  too rapidly with $z$. The corresponding condition limiting the strength of the coherent field, which induces variations of $\hat{\varepsilon}(z;\Omega)$, can be found from Eq.~\eqref{field_eq} differentiating it with respect to $z$. Assuming that the driving field has a peak amplitude $\mathcal{E}_0$, we obtain $\mathcal{E}_0 d\ll n^2$. With the frequencies of the ultrabroadband driving field being on the order of $\Gamma_0$ this
restriction can be reformulated as a condition on the squeezing strength $r$,
\begin{align}\label{Eq:condition_Gamma}
r\ll \frac{n\Gamma_0 l	}{c_0}.
\end{align}

\begin{figure}[t]
\centering
\includegraphics[scale=0.4]{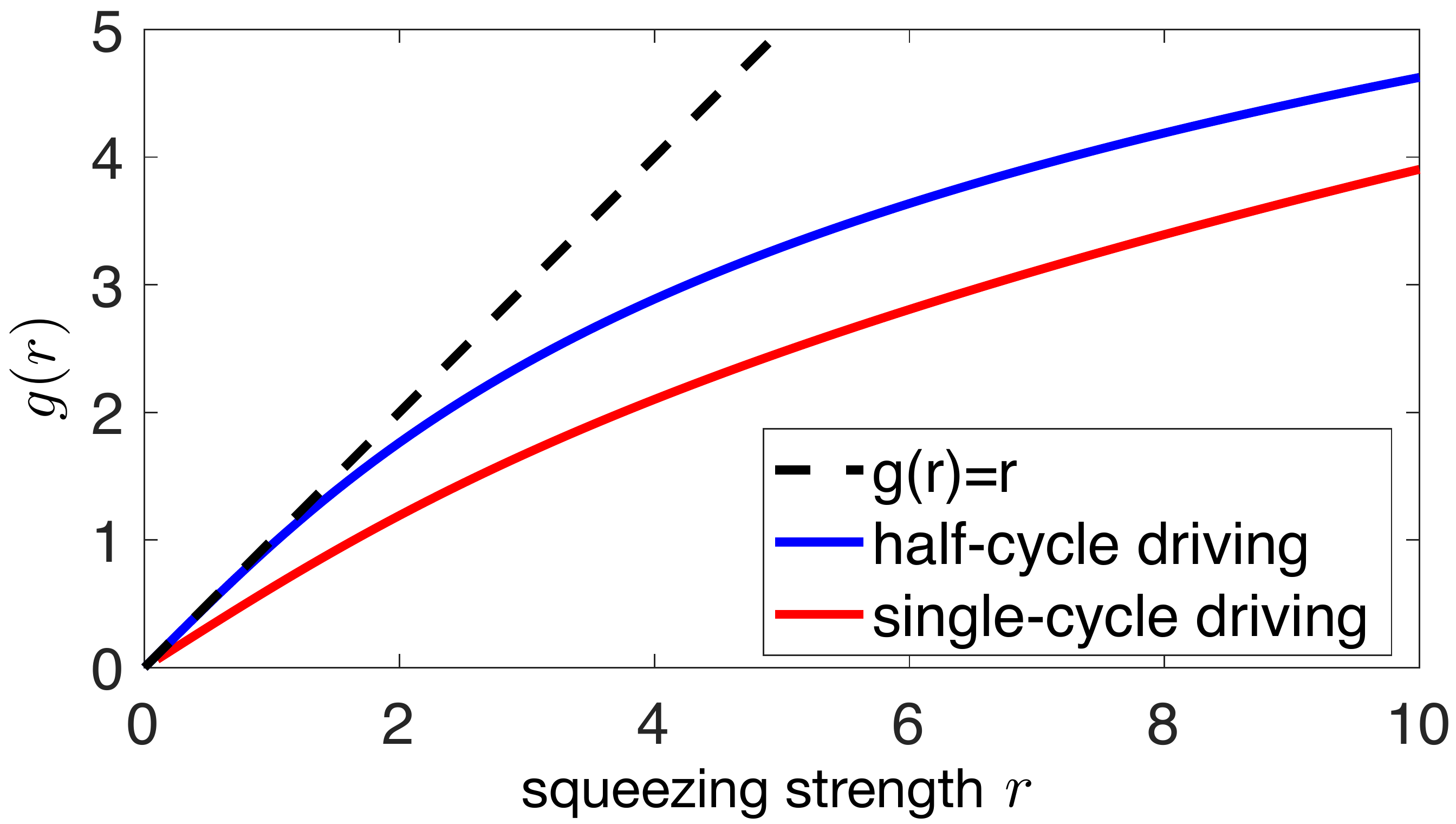}
\caption{ Function $g(r)$
for the half-cycle (blue line) and single-cycle (red line) driving. For $r\ll 1$, $g(r)$
approximately equals $r$ ($0.64r$)
in the case of half-cycle (single-cycle) driving.
For higher squeezing strengths, $g(r)$ grows more slowly with $r$. \label{S2}}
\end{figure}

Solutions obtained within the broadband SVAA do not automatically preserve causality, i.e. the impossibility of information exchange faster than the speed of light. The condition 
\begin{align}
g(r):= \Gamma_0\,\max_{t'}\left[\tau'(z'=l/2,t')-t')\right]\leq \frac{\Gamma_0(n-1)l}{c_0},\label{Eq:condition_Causality}
\end{align}
formulated in the retarded reference frame, ensures causality.
This also effectively sets a boundary for $r$. For $r\ll 1$, the left hand side of the expression above is proportional to $r$ with a proportionality factor given by $\mathrm{max}\,\mathcal{E}'(t')/\mathcal{E}_0\lesssim1$. For higher $r$, the increase becomes even slower [see Fig. \ref{S2}].
Since for nonlinear crystals typically $n$ and $n-1$ are on the same order,
the limit imposed by the range of validity of the broadband SVAA, Eq.~\eqref{Eq:condition_Gamma}, already ensures causality.
This means that it is sufficient to respect Eq.~\eqref{Eq:condition_Gamma}. For squeezing strengths beyond the SVAA the changes in the refractive index induced by the driving field would be comparable to the initial refractive index $n$ of the crystal. Furthermore, effects of higher order nonlinearities like $\chi^{(3)}$ would need to be taken into account.

Considering an experimentally realistic situation with $\Gamma_0/(2\pi)=26$~THz and a $l=15$~$\mu$m thick ZnTe crystal with $n=2.57$, Eq.~\eqref{Eq:condition_Gamma} leads to an estimation
\begin{align*}
r \ll 20.
\end{align*}
In comparison, the limit given by causality in this case can be calculated to
\begin{align*}
g(r)\leq 13.
\end{align*}
In this work the refractive index of the nonlinear crystal is treated as a dispersionless constant. In doing so we neglect the phonon resonances, located for ZnTe at around $5$~THz, and assume that the main contributions to the signal will be given by higher frequencies.
In principle, from Eq.~\eqref{Eq:condition_Gamma} we see that higher values of $r$ become possible if the crystal thickness is increased. This possibility is, however, limited by the requirement to fulfill the ultrabroadband phase matching conditions. Overcoming this issue would open a way for ultrabroadband squeezing in the high gain regime.

\subsection*{Half-cycle pulse}
\begin{figure}[t]
\centering
\includegraphics[scale=0.4]{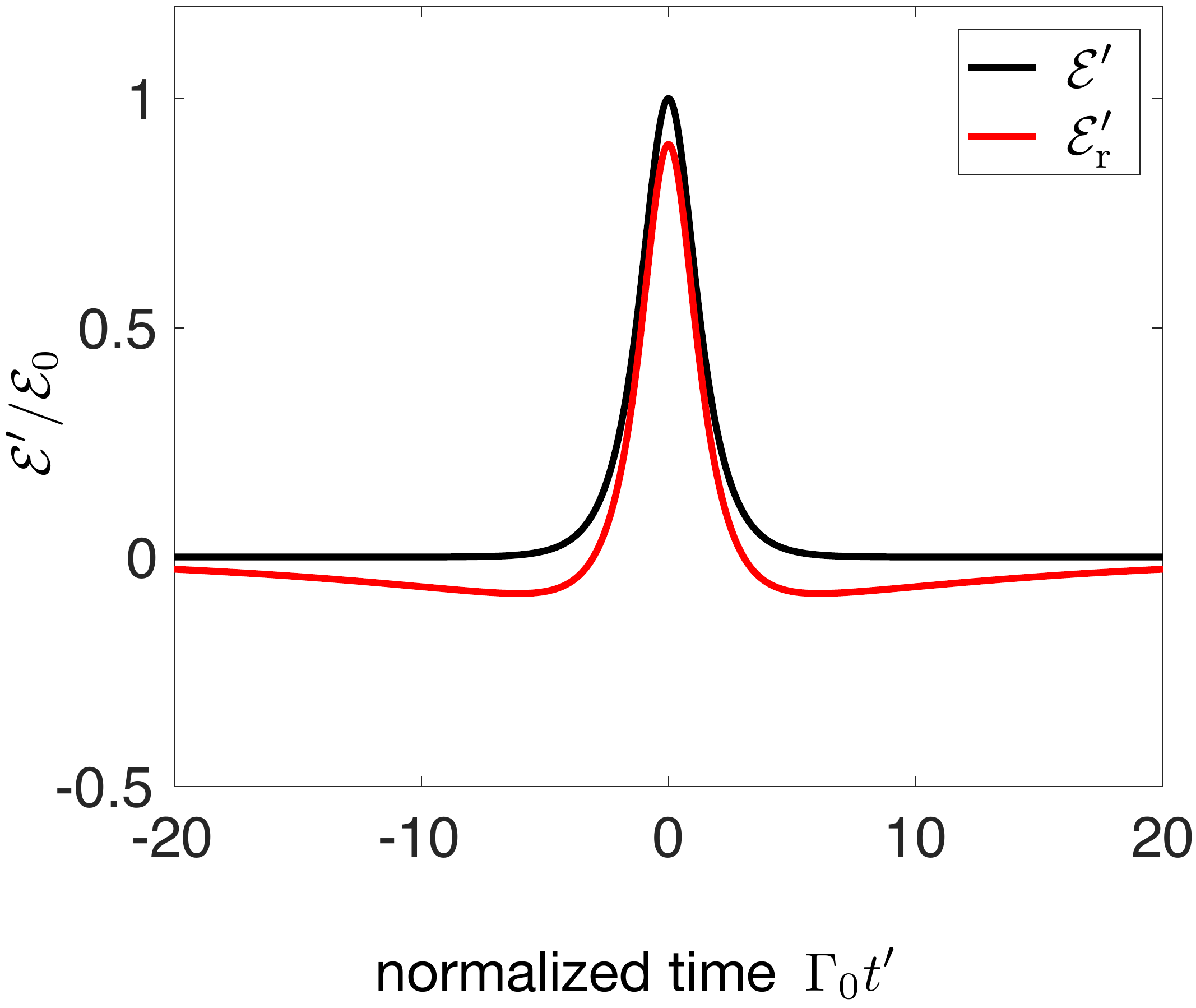}
\caption{Comparison between the half-cycle pulse (black) used for the calculations depicted in Fig. \ref{Fig2} and the more realistic propagating pulse (red) that it approximates. The two side wings of the red curve compensate the big maximum in the center but are still small compared to the maximum.\label{S1}}
\end{figure}

The shape of the half-cycle pulse $\mathcal{E}'(t')=\mathcal{E}_0\sech(\Gamma_0 t')$ used as the driving field in Fig.~\ref{Fig2}
is depicted in Fig.~\ref{S1} as the black line. Because of its non-vanishing integral over time, these type of pulses are
actually not supported in the far-field region with respect to the light source. However, for our particular problem the selected  temporal  profile can serve as a good model of a waveform capable to propagate in the far-field zone, which is relevant to the experimental situation. Let us consider the following allowed pulse shape (red line in  Fig.~\ref{S1})
\begin{align}\label{Eq:E_real}
\mathcal{E}'_\mathrm{r}(t')=\mathcal{E}_0\sech(\Gamma_0 t')-0.1\mathcal{E}_0\sech\left(\frac{\Gamma_0 t'}{10}\right).
\end{align}
The correction introduced by the second term forces the integral over time to vanish.
In principle, one might want to use such or similar shapes for our calculations directly in place of a simpler profile $\mathcal{E}'(t')$. However, the latter has the advantage that the solution for the evolution of the conformal time inside the crystal could be found in an analytical form also for stronger driving fields. Compared with $\mathcal{E}'(t')$, the field $\mathcal{E}'_\mathrm{r}(t')$ has a somewhat lower central positive peak and possesses negative side wings, which are  low-amplitude and extended in time . An important point is that the difference between both discussed waveforms remains small for any moment in time.
Furthermore, in the calculations of the detected variance the difference is even less than it could be anticipated looking at Fig. \ref{S1}.
Indeed, let us consider for example the case of the vanishing probe pulse duration $t_\mathrm{p}$ and weak driving when the dynamics of the detected variance is governed by $\mathrm{d}^3 \mathcal{E}'(t_\mathrm{d})/\mathrm{d} t_\mathrm{d}^3$ or $\mathrm{d}^3 \mathcal{E}'_\mathrm{r}(t_\mathrm{d})/\mathrm{d} t_\mathrm{d}^3$, respectively. In the third derivative the contribution of the second term on the right hand side of Eq.~\eqref{Eq:E_real} is suppressed by an additional factor of $10^3$ with respect to the first term. Thus the difference in the calculated variances may be neglected with a high degree of accuracy.
\subsection*{Fitting procedure for the extraction of the degrees of squeezing}
The red line in Fig.~\ref{Fig3}b depicts the squeezing pattern in the simplified picture.  Here the degree of squeezing is determined by a calculation similar to Eq.~\eqref{Def}, with a minus prefactor (following the convention to assign positive values of the degree of squeezing for squeezing and negative ones for anti-squeezing) and using the ratio between the variances of the outgoing field and vacuum  field taken at the relevant moment. At times where the driving field vanishes this ratio depends exponentially on the squeezing strength $r$.
For such a point inside a squeezing (anti-squeezing) region, it exponentially decays (grows) with $r$. In this case it is guaranteed that the squeezing turns into  anti-squeezing and vice versa by changing the polarity of the driving field. This exponential dependence is shown in Fig.~\ref{Fig3}c as the red lines, where the degree of squeezing at $t_\mathrm{d}=0$ fs is plotted for both polarities of the driving field amplitude in dependence of the squeezing strength $r$.
The observed asymmetry between the maximal positive values of the degree of squeezing (corresponding to squeezing) and its minimal negative values (corresponding to anti-squeezing) can serve as a direct measure of the degree of squeezing.

Figure \ref{Fig3}c demonstrates that a similar measure can be used to extract the degree of squeezing from the RDV within the full description accounting for a finite detection window. The two grey lines reflect the corresponding maximal squeezing and anti-squeezing values obtained for the RDV for different polarities of the driving field amplitude. In order to extract the degree of squeezing from these lines an exponential fit
\begin{equation}\label{Fit}
h_\pm(r)=A_1\big[\exp\left(\pm A_2r\right)-1\big]
\end{equation}
is applied. The degree of squeezing is then given by $-\mathrm{RDV}/A_1$. Since the two fitting functions only differ in the sign of the exponent, the parameters $A_1$ and $A_2$ can be fitted well either to the squeezing or the anti-squeezing values, where the respective differences decrease for shorter probe pulse durations. One should select rather the squeezing branch, since the corresponding dynamics is better resolved than for the anti-squeezing, due to the slower flow of the conformal time in the former case. The result of this fit is shown in Fig.~\ref{Fig3}c as the dashed red lines. For the case of a much shorter probe pulse duration the extracted squeezing and anti-squeezing curves would coincide with the red curves of the simplified picture.

The right axes of Figs.~\ref{Fig3}b and \ref{Fig3}c show the degrees of squeezing obtained in this way from the RDV and within the simplified picture.

\subsection*{Single-cycle RDV for different squeezing strengths}
\begin{figure}[t]
\centering
\includegraphics[scale=0.5]{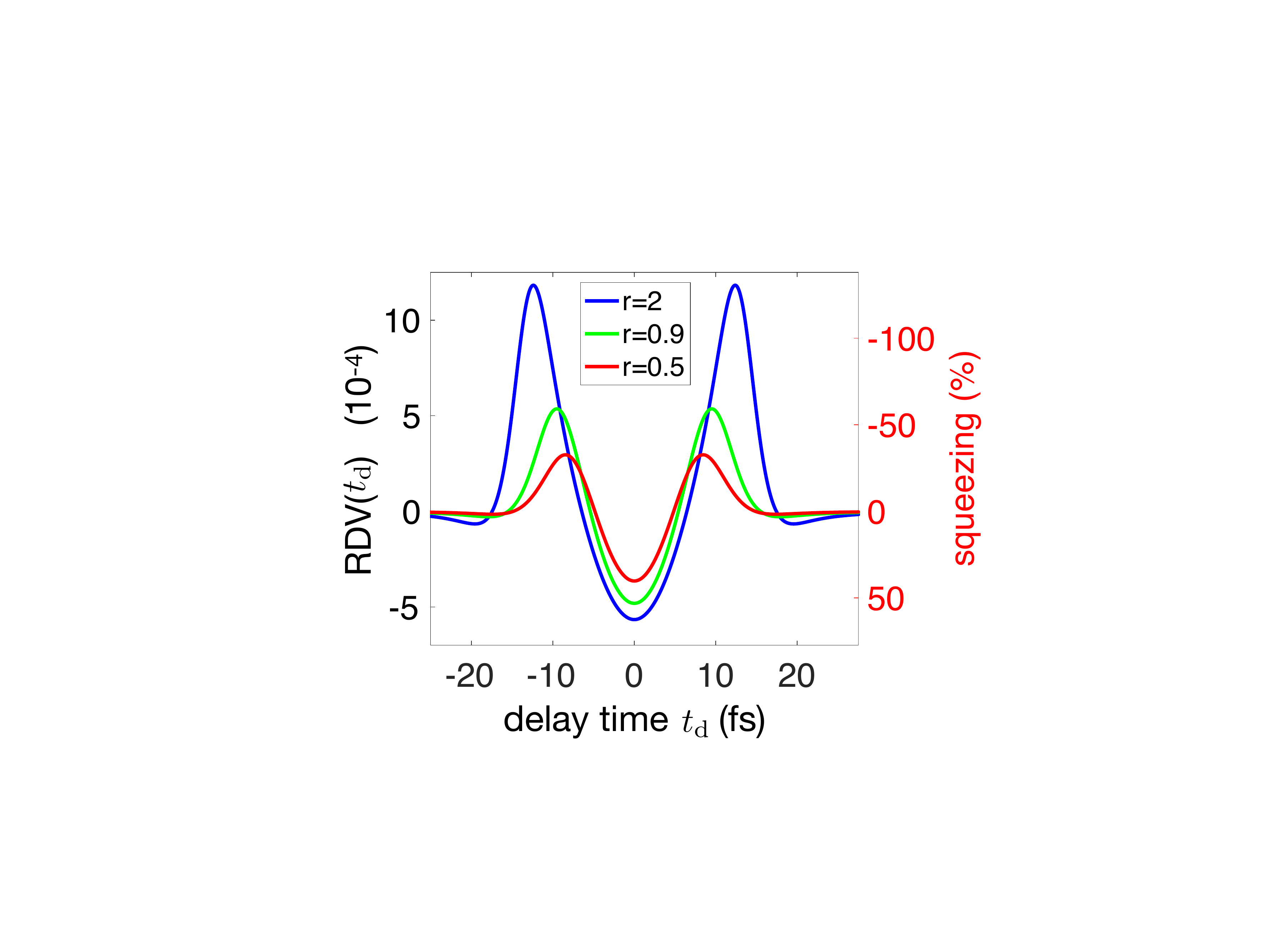}
\caption{Comparison of the RDV for the single-cycle pulse, sampled with a probe pulse with $t_\mathrm{p}=0.49$ fs, for different squeezing strengths.\label{S3}}
\end{figure}

Fig. \ref{S3} depicts the RDV for the single-cycle driving field for different squeezing strengths. For $r=0.5$ the maximal absolute degree of squeezing exceeds the maximal absolute degree of anti-squeezing. Increasing the squeezing strength however restores the conventionally expected asymmetry of more pronounced anti-squeezing than squeezing.

\subsection*{Squeezing operator approach}
It can be shown\cite{Guedes2018} that up to the first order in the driving coherent field
Eq.~\eqref{field_eq} results in the following transformation of the annihilation operator for the quantum field $\hat{\varepsilon}$:
\begin{equation}
\hat{a}(z;\Omega)= \,\hat{a}(\Omega)+\left(\frac{z}{l}+\frac{1}{2}\right)\text{sign} (\Omega)\int^{\infty}_{-\infty}d\Omega_1 \Xi_\mathrm{sym}(\Omega,\Omega_1)\hat{a}^{\dagger}(\Omega_1),
\label{transform_a}
\end{equation}
where $\Xi_\mathrm{sym}(\Omega,\Omega_1)=-\frac{idl}{n c_0} \text{sign}(\Omega\Omega_1)\sqrt{|\Omega\Omega_1|} \mathcal{E}_{\mathrm{\Omega_1+\Omega}}$, $\mathcal{E}_{\Omega}$ denotes the Fourier transform of the driving field in the retarded reference frame $\mathcal{E}'(t')$, $z\in[-l/2,l/2]$ and $\hat{a}(\Omega)\equiv \hat{a}(z=-l/2;\Omega) \equiv \hat{a}_{\mathrm{in}}(\Omega)$ corresponds to the incoming field $\hat{\varepsilon}_\mathrm{in}$. For the outgoing field $\hat{\varepsilon}_\mathrm{out}$ at the end of the crystal we have $\hat{a}(z=l/2;\Omega) \equiv \hat{a}_{\mathrm{out}}(\Omega)$. The transformation of the creation operator is given by the Hermitian conjugate of \eqref{transform_a}.

The linear relations between $\hat{a}_{\mathrm{out}}(\Omega), \hat{a}^\dagger_{\mathrm{out}}(\Omega)$ and $\hat{a}_{\mathrm{in}}(\Omega), \hat{a}^\dagger_{\mathrm{in}}(\Omega)$ can be viewed as a generalized  Bogolyubov transformation \cite{Yablonovitch1989}. It can be induced by a unitary squeezing operator $\hat{S}$ as $\hat{a}_{\mathrm{out}}(\Omega)=
\hat{S}\hat{a}_{\mathrm{in}}(\Omega)\hat{S}^\dagger$ \big[$\hat{a}^\dagger_{\mathrm{out}}(\Omega)=
\hat{S}\hat{a}^\dagger_{\mathrm{in}}(\Omega)\hat{S}^\dagger$\big]. One finds\cite{Guedes2018}
\begin{align}
\hat{S}=\exp\bigg\{\frac{1}{2}\iint^\infty_{0}d\Omega_1d\Omega_2\Big[\Xi^*_\mathrm{sym}(\Omega_1,\Omega_2)\hat{a}_{}(\Omega_1)\hat{a}_{}(\Omega_2)-\Xi_\mathrm{sym}(\Omega_1,\Omega_2)\hat{a}^\dagger_{}(\Omega_1)\hat{a}^\dagger_{}(\Omega_2)\nonumber\\
+\Xi^*_\mathrm{sym}(-\Omega_1,\Omega_2)\hat{a}^\dagger_{}(\Omega_1)\hat{a}_{}(\Omega_2)-\Xi_\mathrm{sym}(-\Omega_1,\Omega_2)\hat{a}_{}(\Omega_1)\hat{a}^\dagger_{}(\Omega_2)\Big]\bigg\}.
\label{squeezing}
\end{align}
This continuous-mode form of the operator $\hat{S}$ extends \cite{Blow1990} the definition of a discrete multimode squeezing operator \cite{Vogel_book,Lo1993}. Moreover, since in a thin crystal the generated photon pairs copropagate along the same axis, both frequency conversion and parametric down conversion \cite{Christ2013} happen simultaneously.
This leads to the creation of photons in an even broader frequency range.
The resulting broadband quantum state of light in the Schr\"{o}dinger picture is obtained by applying the squeezing operator \eqref{squeezing} to the vacuum state $|0\rangle$: $|\{\xi_\Omega\}\rangle=\hat{S}|0\rangle$.

The outgoing quantum field in the retarded reference frame reads
\begin{equation}
\hat{\varepsilon}'_\mathrm{out}(t') =i\int^\infty_{0}d\Omega\,\sqrt{\frac{\hbar \Omega}{4\pi n\epsilon_0 c_0 A}}\left[\hat{a}_\mathrm{out}(\Omega)e^{-i\Omega t'} - \hat{a}_\mathrm{out}^\dagger(\Omega)e^{i\Omega t'}\right],
\label{quad_E}
\end{equation}
where $A$ is the normalization area and we have taken into account the relation $k_\Omega=\Omega n/c_0$. In this work we are interested in the difference in the detected variances for the outgoing and incoming fields $\big\langle 0\big|[\hat{\varepsilon}^{\mathrm{(d)}}_{\mathrm{out}}(t_\mathrm{d})]^2\big|0\big\rangle
-\big\langle 0\big|[\hat{\varepsilon}^{\mathrm{(d)}}_{\mathrm{in}}]^2\big|0\big\rangle$, which occurs in Eq.~\eqref{Def}.
For the incoming vacuum field the detected variance does not depend on the delay time $t_\mathrm{d}$ of the probe field. In the limit case of the vanishing probe duration $t_\mathrm{p}$, the considered difference converges to the normally ordered \cite{Vogel_book} variance of the outgoing quantum field
$\langle 0|\normord{[\hat{\varepsilon}_{\mathrm{out}}(t_\mathrm{d})]^2}| 0\rangle$ at the time $t=t_\mathrm{d}$. Using  Eq.~\eqref{transform_a} and Eq.~\eqref{quad_E} we can compute this quantity up to the first order in the driving field. In principle, it is possible to obtain also the higher order terms \cite{Guedes2018}. For that, the further expansion terms should be calculated in Eq.~\eqref{transform_a} from the non-perturbative result for the operator $\hat{S}$ provided by Eq.~\eqref{squeezing}.

Let us consider a driving field with a half-cycle shape \eqref{HCP}. Its Fourier transform reads $\mathcal{E}_\Omega=\frac{\mathcal{E}_0}{2\Gamma_0} \text{sech}(\frac{\pi\Omega}{2\Gamma_0})$. Then up to the first order in perturbation theory we can calculate
\begin{equation}
\langle 0|\normord{[\hat{\varepsilon}_{\mathrm{out}}(t_\mathrm{d})]^2}| 0\rangle = \frac{\hbar \Gamma^2_0}{4\pi n\epsilon_0 c_0 A}\frac{r}{6}\left[\text{tanh}^3(\Gamma_0 t_\mathrm{d})\text{sech}(\Gamma_0 t_\mathrm{d})-5\text{tanh}(\Gamma_0 t_\mathrm{d})\text{sech}^3(\Gamma_0 t_\mathrm{d}) \right].
\end{equation}
This result holds when $r=|\mathcal{E}_0d|\Gamma_0 l/(nc_0) \ll 1$. It can be seen that under these conditions the variance is an odd function of time (cf. dashed black line in Fig.~\ref{Fig2}b) and scales linearly with the squeezing strength $r$. Thus the temporal profiles of the squeezing and anti-squeezing remain symmetric with respect to each other until the limit of validity of the utilized approximation is reached.

\bibliography{report}

\end{document}